\documentclass[a4paper]{PoS}
\usepackage{amsmath}
\usepackage[font=small,labelfont=bf]{caption}

\title{NLO predictions for SMEFT in the top-quark sector}

\ShortTitle{NLO SMEFT for top}

\author{\speaker{Cen Zhang}\\
        Department of Physics, Brookhaven National Laboratory\\
	Upton, NY 11973, USA\\
        E-mail: \email{cenzhang@bnl.gov}}

\abstract{
	Predictions for the Standard Model Effective Field Theory at the
	next-to-leading order accuracy in QCD, including parton-shower effects,
	have started to become available in the {\sc MadGraph5\_aMC@NLO}
	framework.  In this talk we summarize some recent
	results for $t\bar t$, single top, $t\bar tZ/\gamma$, and $t\bar tH$
	production channels at dimension six.
}

\FullConference{38th International Conference on High Energy Physics\\
		3-10 August 2016\\
		Chicago, USA}

\begin{document}

\section{Introduction}
The Standard Model Effective Field Theory (SMEFT) at
dimension-six \cite{Buchmuller:1985jz} is a powerful approach to
the SM deviations. By
supplementing the SM Lagrangian with a set of higher-dimensional operators,
indirect effects from heavy particles, possibly beyond the reach of the LHC, can
be consistently accommodated.   Given that the expectations from LHC Run-II on
the attainable precision of the top-quark measurements are very high,
next-to-leading order (NLO) predictions for top-quark production channels 
are becoming relevant, not only for the SM background but also for the deviations
from dimension-six operators, mainly for the following reasons:
\begin{itemize}
	\item At the LHC, the impact of QCD corrections on total
		cross sections are often
		large, which might improve the exclusion limits on
		effective operators.  In addition, NLO corrections reduce the
		theoretical uncertainties due to missing higher-order
		corrections. This helps to discriminate between different new
		physics scenarios.
	\item QCD corrections often change the distributions of key observables.
		As
		differential distributions start to play an important role in
		recent global analyses based on SMEFT, reliable predictions for the
		distributions are needed.  In section \ref{sec:singletop} we
		will show an example where this effect is crucial.

		\iffalse
	\item The SMEFT has many operator coefficients, and many of them
		remain to be constrained. Thus higher order effects are important
		when unconstrained or loosely-constrained operators enter
		at one loop (see, for example,
		\cite{Degrande:2012gr,Greiner:2011tt,Zhang:2012cd,
		Mebane:2013cra,Mebane:2013zga,deBlas:2015aea,Elias-Miro:2013eta}).

	\item NLO is important to understand the structure of the
		effective theory, mainly because going to higher order
		allows us to control the effects of renormalization group
		(RG) and operator mixing
		\cite{Jenkins:2013zja,Jenkins:2013wua,Alonso:2013hga}, and the
		corresponding theoretical uncertainties due to missing
		higher-order terms. 
		\fi

	\item Sensitivity to effective
		deviations can be improved by making use of the accurate SMEFT
		predictions and designing optimized experimental strategies in
		a top-down way.  However, given the large QCD corrections at
		the LHC, this improvement will be difficult without consistent
		SMEFT at NLO predictions. 
\end{itemize}
%The above points motivate us to study the higher-order corrections within
%the SMEFT, and in particular, the NLO QCD corrections
%are usually the most important ones at the LHC.

Recently, NLO predictions for the SMEFT, matched with parton shower simulation, are
becoming available in the {\sc MadGraph5\_aMC@NLO} framework
\cite{Alwall:2014hca}, based on an automatic approach to NLO QCD calculation
interfaced
with shower via the {\sc MC@NLO} method \cite{Frixione:2002ik}.  The dimension-six
Lagrangian can be implemented with the help of a series of packages,
including {\sc FeynRules} and {\sc NLOCT} \cite{Alloul:2013bka, Degrande:2014vpa,
Degrande:2011ua,deAquino:2011ub,Hirschi:2011pa,Frederix:2009yq}.
A model in the Universal {\sc FeynRules} Output format \cite{Degrande:2011ua}
can be built, allowing for simulating a variety of processes at NLO in QCD. In
this talk we summarize some recent progresses in this direction, with a focus
on the top-quark sector. The interested readers may find more details in
Refs.~\cite{Franzosi:2015osa,Zhang:2016omx,Bylund:2016phk,Maltoni:2016yxb}.

\section{Top-pair production}
\begin{figure}[ht]
 \begin{minipage}{\textwidth}
  \begin{minipage}[b]{0.49\textwidth}
	  \centering
		\includegraphics[width=\linewidth]{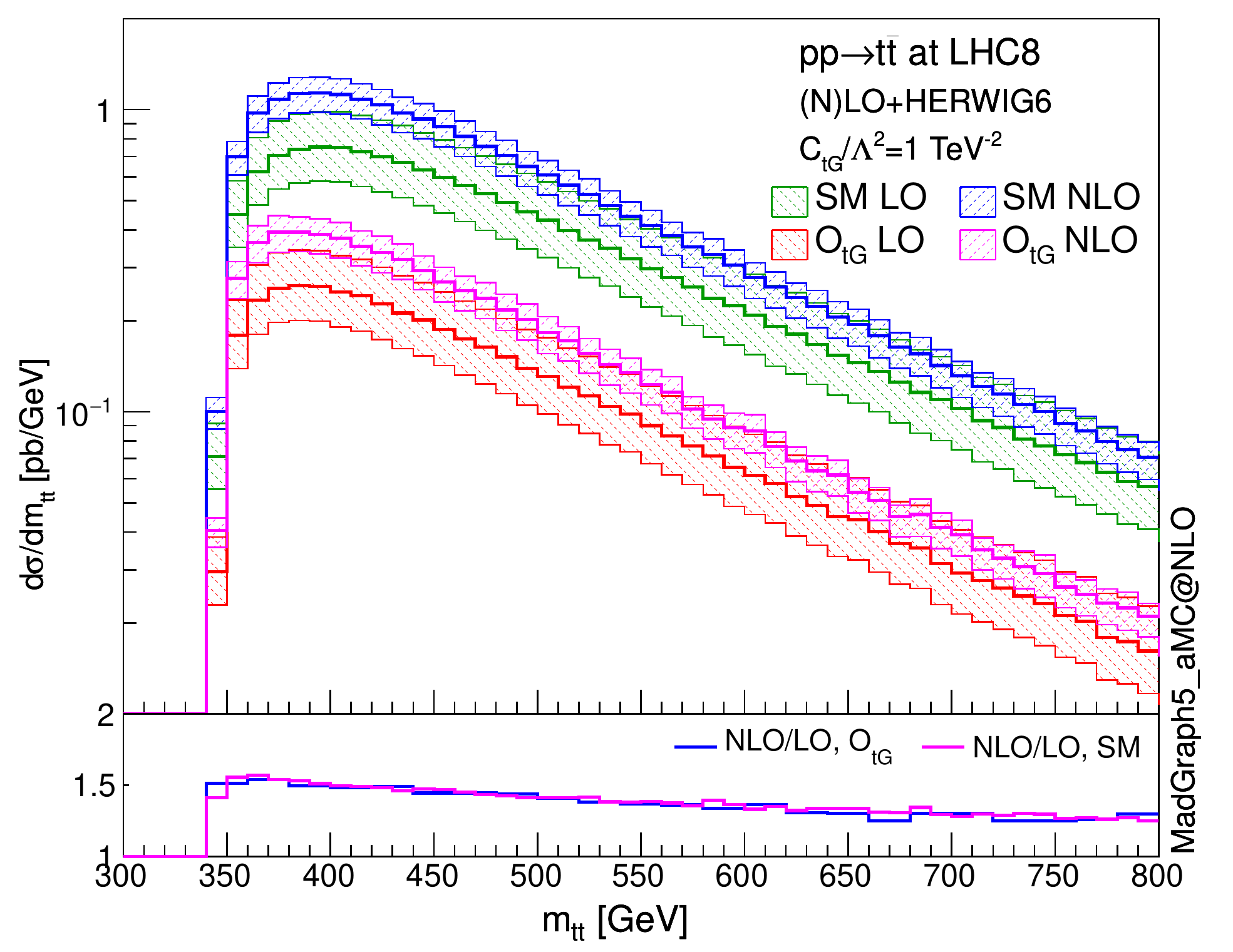}
		\captionof{figure}{Top quark pair invariant mass distribution at LHC 8 TeV.\label{fig:tt}}
  \end{minipage}
  \hfill
  \begin{minipage}[b]{0.49\textwidth}
	  \centering
\begin{tabular}{|l|c|c|}
\hline
& LO [TeV$^{-2}$]  & NLO [TeV$^{-2}$] \\
\hline
Tevatron& [-0.33, 0.75]	   &  [-0.32, 0.73]	\\
\hline
LHC8& [-0.56, 0.41]  &  [-0.42, 0.30]	\\
\hline
LHC14& [-0.56, 0.61]     &  [-0.39, 0.43]\\
\hline
\end{tabular}
  \captionof{table}{Limits on $C_{tG}/\Lambda^2$. 
The corresponding limits 
combining Tevatrion and LHC8, in terms of
$d_V$, is $[-0.0099,0.0123]$ at LO and $[-0.0096,0.0090]$ at NLO.
For LHC14 we assume a $5\%$ experimental error.
\label{tab:tt}
}
	\end{minipage}
    \end{minipage}
\end{figure}

The chromo-dipole operator for the top quark,
$	O_{tG}=g_sy_t\left(\bar Q\sigma^{\mu\nu}T^At\right)\tilde\phi G^A_{\mu\nu}$,
can be constrained by top-pair production. 
Here $g_{s}$ is the strong interaction coupling, and $y_t$ is the top Yukawa
coupling. $Q$ is the third generation left-handed quark doublet, while
$t$ is the right-handed top quark.  Assuming real operator coefficient, this
calculation has been carried out at NLO in Ref.~\cite{Franzosi:2015osa}.
In Figure~\ref{fig:tt} we present the invariant mass distribution at LHC 8 TeV.
The $K$-factors for the total cross sections are found to be $1.1$, $1.4$, and $1.5$
respectively for Tevatron, LHC 8 TeV, and LHC 13/14 TeV. As a result, the current
limits on the chromomagnetic dipole moment of the top quark from direct measurements
can be improved by roughly the same factors.  In Table~\ref{tab:tt} we compare
the limits on $C_{tG}/\Lambda^2$ at LO and at NLO.

\section{Single top production}\label{sec:singletop}
\begin{figure}[ht]
 \begin{minipage}{\textwidth}
  \begin{minipage}[b]{0.49\textwidth}
	  \centering
		\includegraphics[width=\linewidth]{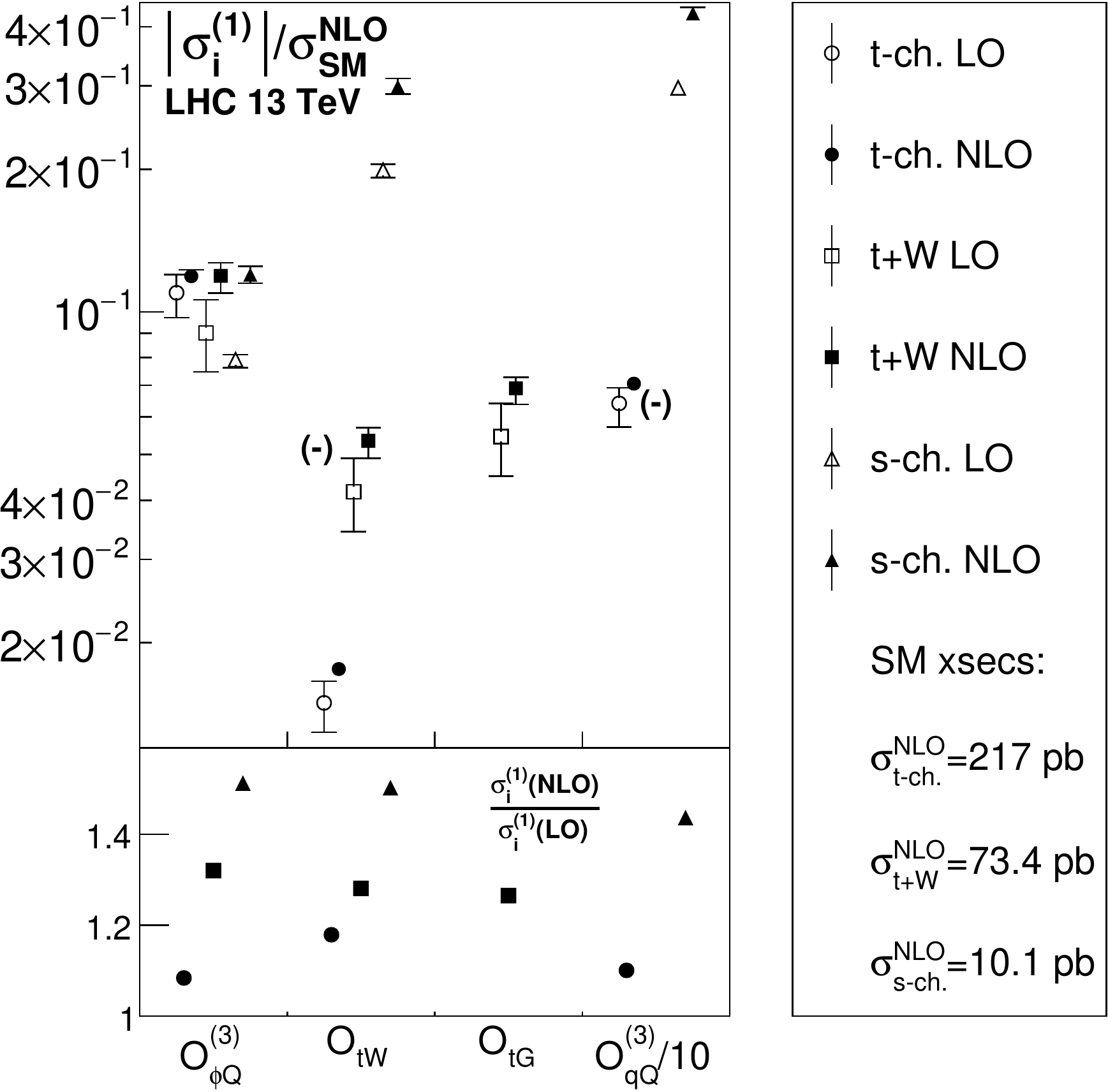}
	\caption{$r_i=\left|\sigma^{(1)}_i\right|/\sigma_{\mathrm{SM}}^{\mathrm{NLO}}$
	for the three single-top channels.  Both LO and NLO results are shown.
	Error bars indicate scale uncertainties.  $K$-factors are given in the
	lower panel.  Negative contributions are labeled with ``(-)''.
	\label{fig:t}}
  \end{minipage}
  \hfill
  \begin{minipage}[b]{0.49\textwidth}
	  \centering
		\includegraphics[width=\linewidth]{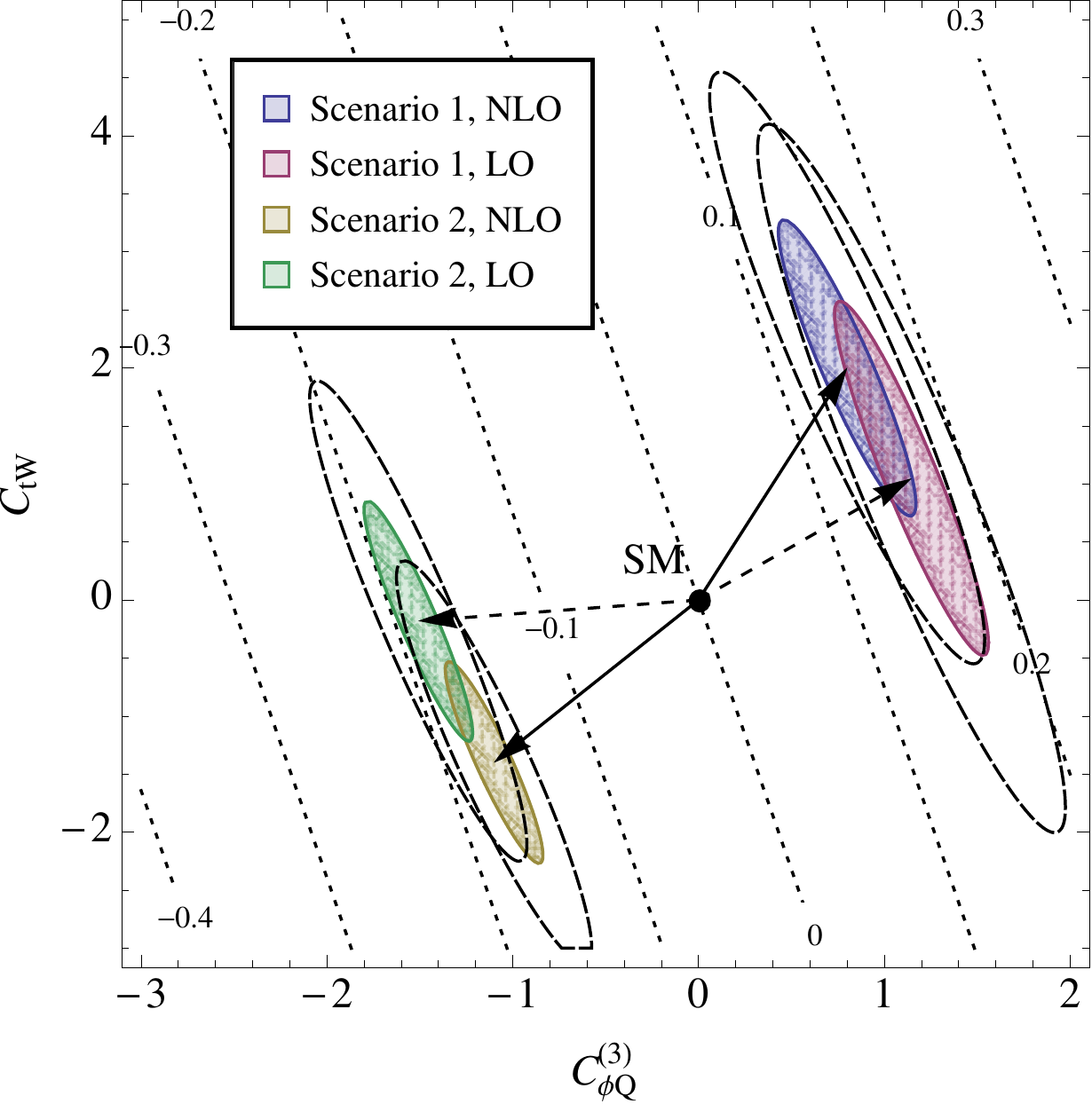}
	\caption{Two-operator fit using single-top pseudomeasurements on
	shapes, at 68\% confidence level, assuming 5\% uncertainty in each bin.
Dashed lines correspond to twice this uncertainty, while dotted contours are
the relative deviation in total cross section.}
		\label{fig:shapefit}
	\end{minipage}
    \end{minipage}
\end{figure}

Single top production has been computed in all three channels ($t$-channel,
$s$-channel, and $tW$ associated production channel) at NLO in QCD, with the following
operators:
\newcommand{\FDFI}{\left(\phi^\dagger\overleftrightarrow{D}^I_\mu\phi\right)}
\begin{flalign}
  &O_{\phi Q}^{(3)}
  =i\frac{1}{2}y_t^2 \FDFI (\bar{Q}\gamma^\mu\tau^I Q)\;,
  \qquad
  O_{tW}=y_tg_W(\bar{Q}\sigma^{\mu\nu}\tau^It)\tilde{\phi}W_{\mu\nu}^I\;,
  %&O_{tG}=y_tg_s(\bar{Q}\sigma^{\mu\nu}T^At)\tilde{\phi}G_{\mu\nu}^A\;,
  \\
  &O_{qQ,rs}^{(3)}=(\bar q_r\gamma_\mu\tau^Iq_s)(\bar Q\gamma^\mu\tau^IQ)\;,
  \label{eq:Otf}
\end{flalign}
and $O_{tG}$ \cite{Zhang:2016omx}. Here $q_r$ and $q_s$ are the quark doublet fields in the first two
generations. $r,s$ are flavor indices.  $g_W$ is the SM weak coupling
constant.  The operators $O_{tG}$ and $O_{tW}$ have mixing effect.
Total cross sections (including top and antitop) at LHC 13 TeV are presented
in Figure~\ref{fig:t}.  The ratios between the interference cross
sections, $\sigma^{(1)}_i$, and the SM NLO cross section,
$r_i=\left|\sigma^{(1)}_i\right|/\sigma_{\mathrm{SM}}^{\mathrm{NLO}}$, for
individual operators $O_i$, are given in all three channels. Scale
uncertainties from the numerator are given, and in the lower panel the
$K$-factor of each operator contribution is shown.  Improved accuracy is reflected by
the $K$-factors, typically ranging from $\sim10\%$ to $\sim50\%$, and improved
precision is reflected by the significantly reduced scale uncertainties.
%One can clearly see that NLO results are outside of the uncertainty range of
%corresponding LO results.

QCD corrections to the shapes of discriminator
observables could lead to bias in an SMEFT analysis, by shifting the theoretical
predictions for the shapes of the observables.
This effect might lead to a different direction in which new physics deviates
from the SM. As a result, if deviations due to new physics are observed, missing
QCD corrections could lead us to misinterpret the measurements
and misconclude the nature of UV physics.  An example of a
two-operator fit using pseudomeasurements in $t$-channel single top is given in
Figure~\ref{fig:shapefit}, assuming two scenarios: $(C_{\phi Q}^{(3)},C_{tW})
=(0.8,2.0)$, and $(C_{\phi Q}^{(3)},C_{tW})=(-1.1,-1.4)$. More details can be found in
Ref.~\cite{Zhang:2016omx}.

\section{Top-pair production in association with a gauge boson}

\begin{figure}[ht]
	  \centering
\includegraphics[width=.65\textwidth]{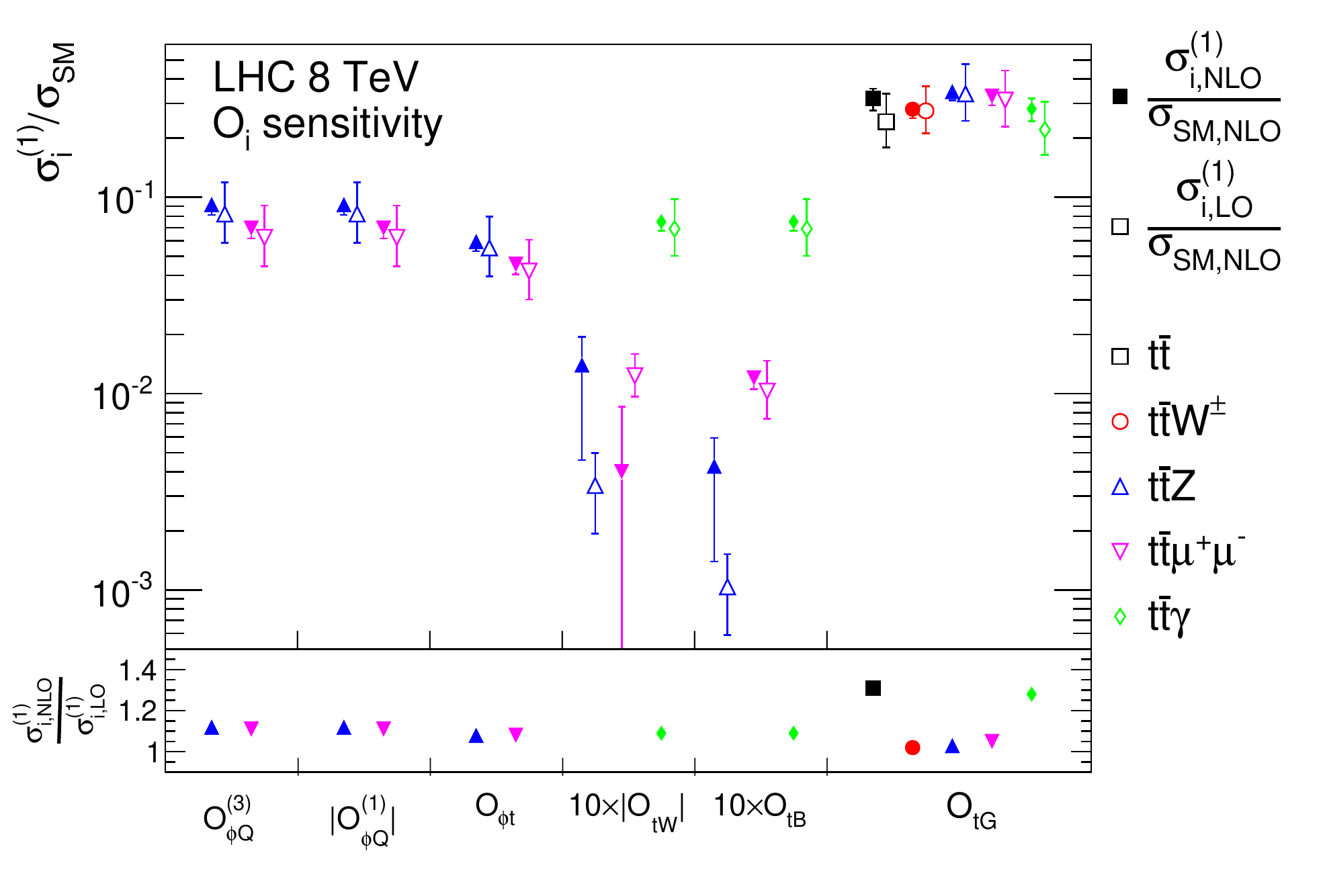}
\caption{Sensitivity of various top quark processes to the various operators
	shown at LO and NLO at 8 TeV. $K$-factors are also shown for
	$\sigma^{(1)}_i$ as well as the scale uncertainties. We do not show the
	$K$-factors for the $O_{tB}$ and $O_{tW}$ operators
	in the $t\bar{t}Z$ and $t\bar{t}\mu^+\mu^-$ processes, as there are
	accidental cancellations that lead to large or even negative $K$-factors.
	\label{fig:ttz}
}
\end{figure}

At the LHC, the neutral couplings $ttZ$ and $tt\gamma$ can be probed
by associated production of a top-quark pair with a neutral gauge boson
$Z/\gamma$.  The relevant operators, apart from $O_{tG}$, $O_{\phi
Q}^{(3)}$, and $O_{tW}$, are:
\newcommand{\FDF}{\left(\phi^\dagger\overleftrightarrow{D}_\mu\phi\right)}
\begin{flalign}
	&O_{\phi Q}^{(1)} =i\frac{1}{2}y_t^2 \FDF
	(\bar{Q}\gamma^\mu Q)\;,\qquad
	O_{\phi t} =i\frac{1}{2}y_t^2 \FDF
	(\bar{t}\gamma^\mu t) \;,
	\\ &O_{tB}=y_tg_Y(\bar{Q}\sigma^{\mu\nu}t)\tilde{\phi}B_{\mu\nu}\;.
\end{flalign}
The corresponding NLO predictions are given in Ref.~\cite{Bylund:2016phk}.
Here we only present a summary plot for total cross sections in
Figure~\ref{fig:ttz}, similar to Figure~\ref{fig:t}.  By studying the
differential distributions, we also find that the differential $K$-factor of
the SM and that of the operator contribution can be quite different, therefore
using the SM $K$-factor to rescale the operator contributions may not be a good
approximation.  Finally, $e^+e^-\to t\bar t$ with the same operators have also
been computed.

\section{Top-pair production in association with a Higgs boson}

The LHC provides us the first chance to directly measure the interactions between
the top quark and the Higgs boson through the associated production of a Higgs with
$t\bar t$.  In Ref.~\cite{Maltoni:2016yxb}, this process has been computed
at NLO including three
operators: the chromo-dipole operator $O_{tG}$, the Yukawa operator
$O_{t\phi}=y_t^3 \left( \phi^\dagger\phi \right)\left( \bar Qt \right)\tilde\phi$, and the Higgs-gluon operator $O_{\phi G}= y_t^2 \left( \phi^\dagger\phi \right) G_{\mu\nu}^A G^{A\mu\nu}$. 
%where
%\begin{flalign}
%	O_{t\phi} = y_t^3 \left( \phi^\dagger\phi \right)\left( \bar Qt \right)
%	\tilde\phi \,,\qquad
%	O_{\phi G} = y_t^2 \left( \phi^\dagger\phi \right) G_{\mu\nu}^A G^{A\mu\nu}\,.
%\end{flalign}
The QCD mixing of these three operators goes in the direction of increasing
number of Higgs fields, i.e.~$O_{tG}$ mixes into $O_{\phi G}$, and both of them
mix into $O_{t \phi}$, but not the other way around.

Cross sections from dimension-six operators can be parametrized as 
\begin{flalign}
	\sigma=\sigma_{SM}+\sum_i\frac{1{\rm TeV}^2}{\Lambda^2}C_i\sigma_i
	+\sum_{i\leq j}
	\frac{1{\rm TeV}^4}{\Lambda^4}C_iC_j\sigma_{ij}.
	\label{eq:xsecpara}
\end{flalign}
As an example we present here $\sigma_{i}$ and $\sigma_{ij}$ in Table~\ref{tab:tth}.
For each central value we quote three uncertainties. The first is the standard scale
uncertainty from renormalization and factorization scales.  The third
uncertainty comes from the PDF sets. The second one is new in the SMEFT.
It comes from the EFT scale uncertainty, representing the missing higher-order
corrections to the operators.
%obtained by using the following quantities
%\begin{flalign}
%	&\sigma_i(\mu_0;\mu)=\Gamma_{ji}(\mu,\mu_0)
%	\sigma_j(\mu)\,,\quad
%	\sigma_{ij}(\mu_0;\mu)= \Gamma_{ki}(\mu,\mu_0) \Gamma_{lj}(\mu,\mu_0)
%	\sigma_{kl}(\mu)\,,
%	\label{eq:sigma2}
%\end{flalign}
%with $\mu_0=m_t$ and $\mu$ varying between $m_t/2$ and $2m_t$ to assess the EFT
%scale uncertainty. 
%Here the $\Gamma_{ij}$ describes the running of operator coefficients,
%i.e.~$C_i(\mu)=\Gamma_{ij}(\mu,\mu_0)C_j(\mu_0)$.
A more detailed discussion can be found in Ref.~\cite{Maltoni:2016yxb}.

\newcommand{\xs}[7]{$#1^{+#2+#6+#4}_{-#3-#7-#5}$}
\begin{figure}[ht]
 \begin{minipage}{\textwidth}
  \begin{minipage}[b]{0.49\textwidth}
	  \small
	  \centering
	\begin{tabular}{llllll}
		\hline\hline
		&
		$\sigma$ NLO & $K$
		\\\hline
$\sigma_{SM}$&\xs{0.507}{0.030}{0.048}{0.007}{0.008}{0.000}{0.000}&1.09\\
$\sigma_{t\phi}$&\xs{-0.062}{0.006}{0.004}{0.001}{0.001}{0.001}{0.001}&1.13\\
$\sigma_{\phi G}$&\xs{0.872}{0.131}{0.123}{0.013}{0.016}{0.037}{0.035}&1.39\\
$\sigma_{tG}$&\xs{0.503}{0.025}{0.046}{0.007}{0.008}{0.001}{0.003}&1.07\\
$\sigma_{t\phi,t\phi}$&\xs{0.0019}{0.0001}{0.0002}{0.0000}{0.0000}{0.0001}{0.0000}&1.17\\
$\sigma_{\phi G,\phi G}$&\xs{1.021}{0.204}{0.178}{0.024}{0.029}{0.096}{0.085}&1.58\\
$\sigma_{tG,tG}$&\xs{0.674}{0.036}{0.067}{0.016}{0.019}{0.004}{0.007}&1.04\\
$\sigma_{t\phi,\phi G}$&\xs{-0.053}{0.008}{0.008}{0.001}{0.001}{0.003}{0.004}&1.42\\
$\sigma_{t\phi,tG}$&\xs{-0.031}{0.003}{0.002}{0.000}{0.000}{0.000}{0.000}&1.10\\
$\sigma_{\phi G,tG}$&\xs{0.859}{0.127}{0.126}{0.017}{0.022}{0.021}{0.020}&1.37\\
\hline\hline
	\end{tabular}
	\captionof{table}{NLO cross sections in pb for $pp\to t\bar tH$ at 13 TeV,
		and corresponding $K$-factors.\label{tab:tth}}
	\end{minipage}
  \hfill
  \begin{minipage}[b]{0.49\textwidth}
	  \centering
		\includegraphics[width=.9\linewidth]{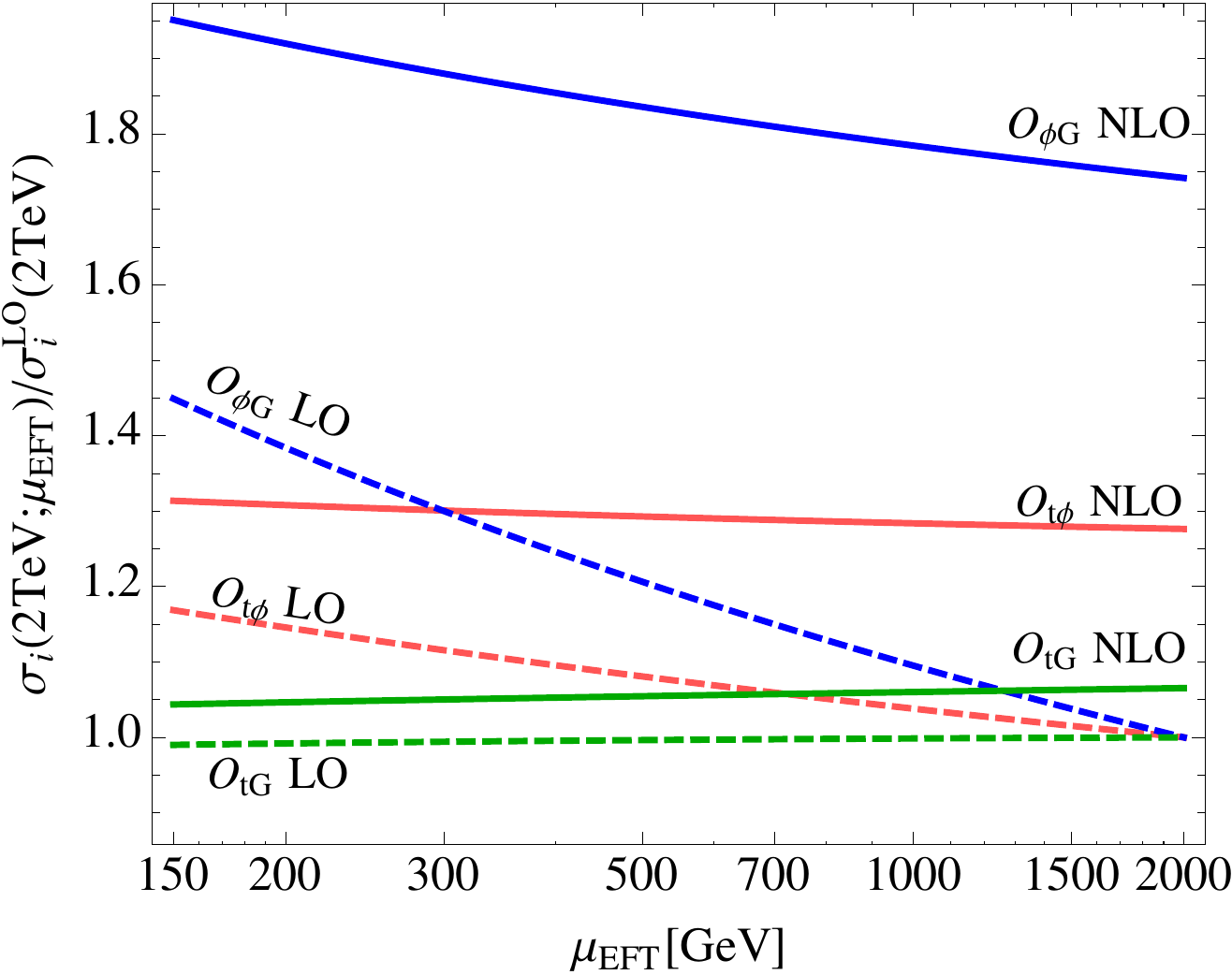}
		\captionof{figure}{Comparison of the RG corrections with the exact NLO results for $t\bar tH$ production.\label{fig:tth}}
  \end{minipage}
    \end{minipage}
\end{figure}
It is also interesting to compare the two kinds of corrections, i.e.~RG and
full NLO, in the $t\bar tH$ process. In Figure~\ref{fig:tth} we show the
interference cross sections from three operators calculated as functions of
$\mu_{EFT}$ for $\Lambda=2$ TeV, where LO contributions are normalized at 2
TeV.
The dashed lines
indicate corrections from one-loop RG only, ranging from
roughly $0$ to $40\%$. Full NLO gives much larger corrections as indicated by
the solid lines.  This plot clearly demonstrates that RG corrections are far from a
good approximation to NLO corrections.
\section{Summary}
\vspace{-.1cm}
We have briefly discussed several recent works on NLO predictions for
SMEFT in the top-quark sector.  These studies pave the way towards an accurate
global fit for top-quark interactions.

\vspace{-.1cm}
\acknowledgments
\vspace{-.1cm}
I would like to thank C.~Degrande, O.~B.~Bylund, D.~B.~Franzosi, F.~Maltoni,
I.~Tsinikos, E.~Vryonidou, and J.~Wang for collaborations on various
top-EFT projects.   The work of
C.Z.~is supported by U.S.~Department of Energy under Grant DE-SC0012704.

\end{document}